\documentclass[fleqn,10pt]{wlscirep}
\usepackage[utf8]{inputenc}
\usepackage[T1]{fontenc}
\usepackage{amsmath}

\title{Nuclear and Magnetic Spin Structure of the Antiferromagnetic Triangular Lattice Compound LiCrTe$_2$ Investigated by $\mu^+$SR as well as Neutron and X-ray Diffraction}

\author[1,*]{E.~Nocerino}
\author[2,3]{C.~Witteveen}
\author[4]{S.~Kobayashi}
\author[5]{O.~K.~Forslund}
\author[1]{N. Matsubara}
\author[6]{A.~Zubayer}
\author[7]{F.~Mazza}
\author[4]{S.~Kawaguchi}
\author[9]{A.~Hoshikawa}
\author[10]{I. Umegaki}
\author[11,12]{J.~Sugiyama}
\author[13]{K.~Yoshimura}
\author[5]{Y.~Sassa}
\author[2]{F.~O.~von~Rohr}
\author[1,$\dagger$]{M.~M{\aa}nsson}

\affil[1]{KTH Royal Institute of Technology, Department of Applied Physics, Alba Nova University Center, Stockholm, SE-114 21, Sweden}
\affil[2]{Department of Quantum Matter Physics, University of Geneva, 24 Quai Ernest-Ansermet
1211 Geneva 4, Switzerland}
\affil[3]{Department of Physics, University of Zürich, Winterthurerstr. 190, 8057 Zürich, Switzerland}
\affil[4]{Japan Synchrotron Radiation Research Institute (JASRI), 1-1-1 Kouto, Sayo 679-5198, Japan}
\affil[5]{Chalmers University of Technology, Department of Physics, G$\ddot{o}$teborg, SE-412 96, Sweden}
\affil[6]{Department of Physics, Chemistry and Biology (IFM), Linköping University, SE-581 83 Linköping, Sweden}
\affil[7]{Insitute of Solid State Physics, TU Wien, Wiedner Haupstraße 8-10, AT-1040 Wien (Austria)}
\affil[9]{Frontier Research Center for Applied Atomic Sciences, Ibaraki University, 162-1 Shirakata, Tokai, Ibaraki 319-1106, Japan}
\affil[10]{Muon Science Laboratory, Institute of Materials Structure Science, KEK, Tokai, Ibaraki 319-1106, Japan}
\affil[11]{Neutron Science and Technology Center, Comprehensive Research Organization for Science and Society (CROSS), Tokai, Ibaraki 319-1106, Japan}
\affil[12]{Advanced Science Research Center, Japan Atomic Energy Agency, Tokai, Ibaraki 319-1195, Japan}
\affil[13]{Department of Chemistry, Graduate School of Science, Kyoto University, Kyoto 606-8502, Japan}

\affil[*]{nocerino@kth.se}
\affil[$\dagger$]{condmat@kth.se}

\begin{abstract}
Two-dimensional (2D) triangular lattices antiferromagnets (2D-TLA) often manifest intriguing physical and technological properties, due to the strong interplay between lattice geometry and electronic properties. The recently synthesized 2-dimensional transition metal dichalcogenide LiCrTe$_2$, being a 2D-TLA, enriched the range of materials which can present such properties. In this work, muon spin rotation ($\mu^+$SR) and neutron powder diffraction (NPD) have been utilized to reveal the true magnetic nature and ground state of LiCrTe$_2$. From high-resolution NPD the magnetic spin order at base-temperature is not, as previously suggested, helical, but rather collinear antiferromagnetic (AFM) with ferromagnetic (FM) spin coupling within the $ab-$plane and AFM coupling along the $c-$axis. The ordered magnetic Cr moment is established as $\mu_{\rm Cr}=2.36~\mu_{\rm B}$. From detailed $\mu^+$SR measurements we observe an AFM ordering temperature $T_{\rm N}\approx125$~K. This value is remarkably higher than the one previously reported by magnetic bulk measurements. From $\mu^+$SR we are able to extract the magnetic order parameter, whose critical exponent allows us to categorize LiCrTe$_2$ in the 3D Heisenberg AFM universality class. Finally, by combining our magnetic studies with high-resolution synchrotron X-ray diffraction (XRD), we find a clear coupling between the nuclear and magnetic spin lattices. This suggests the possibility for a strong magnon–phonon coupling, similar to what has been previously observed in the closely related compound LiCrO$_2$.
\end{abstract}
\begin{document}

\flushbottom
\maketitle
%
%
\thispagestyle{empty}

\section*{Introduction}

For compounds having a two-dimensional triangular lattice with antiferromagnetic interactions (2D-TLA), e.g., the prototypical Na$_{\rm x}$CoO$_2$ \cite{Sugiyama_2003,Hertz_2008,Schulze_2008,Medarde_2013,Sassa_2018}, each corner is occupied by an ion having magnetic moment. When the interaction between neighboring moments is antiferromagnetic (AFM), it is impossible to simultaneously satisfy them all, leading to a geometrical frustration of the system \cite{palmer1984models}. The resulting conflicting atomic interactions often lead to exotic physical properties that can be manifested in many different ways \cite{takada2003superconductivity,mcqueen2008successive,pen1997orbital,katayama2009anomalous,terada2012spiral,arh2022ising}. In this context, chromium compounds of the type $A$Cr$X_2$ ($A$ = monovalent atom, $X$ = chalcogen element) with a triangular lattice (2DTL) are widely studied as geometrically frustrated Heisenberg spin systems with S = 3/2. In particular chromium oxides and sulfides ($X$ = O, S) represent a very popular group of compounds \cite{toth2016electromagnon,ji2009spin,oh2013magnon,seki2008spin,winterberger1987structure,rasch2009magnetoelastic,lafond2001alteration,damay2011magnetoelastic,carlsson2011suppression,Sugiyama_2009}. The iso-structural chromium selenides and tellurides ($X$ = Se, Te) are, on the contrary, much less studied in comparison to their oxides and sulfides counterparts. One of the main reasons is probably that they are difficult to synthesize. However, the few studied members are known to exhibit unexpected structural \cite{Kobayashi_2019} as well as magnetic properties \cite{Sugiyama_2018}, with the formation of very diverse spin structures (e.g. AgCrSe$_2$ exhibits an AFM helical structure while the isostructural NaCrSe$_2$ has ferromagnetic layers stacked antiferromagnetically along the $c$ direction \cite{engelsman1973crystal}). The Li intercalated compound LiCrTe$_2$ represents a valuable addition to this family of materials. Synthesized for the first time only very recently (Kobayashi $et.$ $al.$ 2016 \cite{kobayashi2016competition}), as for the crystal structure databases \cite{wohlfahrt2008springer} and \cite{hellenbrandt2004inorganic}, it presents a 2DTL structure where the Cr$^{3+}$ ions occupy the corners of the triangular lattice and are octahedrally coordinated by Te atoms, as shown by in-house X-ray diffraction at room temperature. Temperature dependent magnetic susceptibility measurements on LiCrTe$_2$ show the typical AFM cusp at $T_{\rm N}\approx71$~K, with a Curie-Weiss temperature $\Theta=58$~K, for an applied magnetic field $H_{\rm ext}=7$~T and $\Theta=101$~K, for $H_{\rm ext}=1$~T \cite{kobayashi2016competition}. The value of Curie-Weiss temperature is related to the strength of the magnetic interactions between ions, and a positive value for $\Theta$ would imply FM correlations \cite{mugiraneza2022tutorial}. In layered materials $\Theta$ usually accounts only for the interactions within the layer, since they are much stronger than the inter-layer ones. Having Cr$^{3+}$ as a magnetic ion, these interactions in LiCrTe$_2$ can be either ferromagnetic (FM), driven by a super-exchange mechanism through an intermediate anion, or AFM, driven by a direct exchange mechanism between the $t_{2g}$ orbitals of two adjacent Cr atoms \cite{kanamori1959superexchange}. The predominance of one mechanism over the other depends largely on the Cr-Cr distance. Given the phenomenology of the magnetic susceptibility measurements, the authors of reference \cite{kobayashi2016competition} suggest that a possible competition of AFM and FM interactions between neighbouring Cr atoms might take place in LiCrTe$_2$, so they conjectured that the spin structure in this material would be helical.

In this work we solved the magnetic structure for polycristalline LiCrTe$_2$ and extracted its temperature dependent magnetic order parameter by means of neutron powder diffraction (NPD), synchrotron X-ray diffraction (XRD) and muon spin rotation ($\mu^+$SR). Our study shows that the FM coupling is dominant within the layer, resulting in a A-type AF structure, with the Cr magnetic moments aligned parallel to each other within the $ab-$plane and anti-parallel along the $c-$axis (i.e. not helical). From the profile of the magnetic order parameter, it emerges that LiCrTe$_2$ can be identified as a 3D Heisenberg AFM.

\section*{Results}

In the following sections the experimental results and detailed analysis are presented. The first subsection shows the temperature dependent $\mu^+$SR measurements in weak transverse field (wTF) and zero field (ZF) configurations, while the second subsection details the synchrotron XRD and NPD measurements, with the respective structural and magnetic refinement for different temperatures.

\subsection*{\label{nopres}$\mu^+$SR Results}

\begin{figure}[ht]
  \begin{center}
    \includegraphics[scale=0.55]{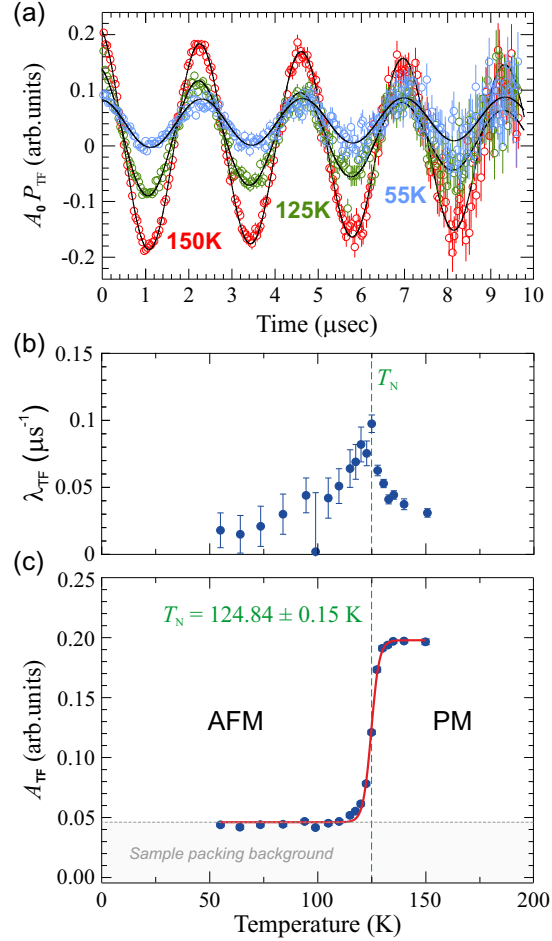}
  \end{center}
  \caption{(a) Weak transverse field (wTF) $\mu^+$SR spectra in the long time domain above the transition temperature for LiCrTe$_2$. The spectra are overlapped to show the oscillation around 0 asymmetry. (b) Temperature dependence of the depolarisation rate of the wTF signal. (c) Temperature dependence of the wTF asymmetry, the solid line is a fit to a sigmoid curve. The transition temperature resulting from the fit is given in the plot.}
  \label{allasy}
\end{figure}

In the wTF geometry an external magnetic field is applied orthogonal to the initial spin-polarization of the muon beam, causing it to precess in the plan perpendicular to the external field´s direction. In this configuration, $\mu^+$SR time spectra have been acquired in the temperature range $T=55-150$~K. The applied transverse field was set to wTF~=~30~G, which is several orders of magnitude smaller than the internal field at the muon sites. Some selected wTF spectra acquired are shown in Fig.~\ref{allasy}(a) as an example of the temperature evolution of the muon spin precession in this set up.

At high temperature the magnetic moments in the sample are randomly oriented, therefore the local internal magnetic field is negligible with respect to the external applied field, which causes the muon spins to precess accordingly. Indeed, the spectrum at $T=150$~K exhibits a very regular oscillation, with a single frequency related to the wTF; then, as the temperature decreases, the system gradually evolves from the paramagnetic (PM) phase to the AFM ordered phase. When entering the AFM state, the internal field compete with the weak external wTF, hereby increasing the overall depolarization rate and consequently reducing the amplitude of the wTF oscillation in the spectra. Finally, only the high frequency oscillations in the early time domain are left. The fit function used for the wTF spectra is the following:

\begin{eqnarray}
 A_0 \, P_{\rm TF}(t) = A_{\rm TF}\cos(2\pi f_{\rm TF}t + \frac{\pi \phi}{180})\cdot{}e^{(-\lambda_{\rm TF} t)} + A_{\rm BG}\cdot{}e^{(-\lambda_{\rm BG} t)}
\label{4}
\end{eqnarray}

Here $A_0$ is the initial asymmetry, $P_{\rm TF}$(t) is the muon spin polarization function, $A_{\rm TF}$ is the asymmetry of the oscillating part of the signal and $A_{\rm BG}$ is the asymmetry of a background component with a slow depolarization rate $\lambda_{\rm BG}$, $2\pi f_{\rm TF}$ is the angular frequency of the Larmor precession (whose value is related to the applied wTF), $\phi$ is the phase of the oscillating signal, $\lambda_{\rm TF}$ (plotted in Fig.~\ref{allasy}(b)) is the depolarisation rates for the respective polarization component. The transition temperature for the sample can be determined analysing the fitting parameter A$_{\rm TF}$. Figure~\ref{allasy}(c) shows a plot of the transverse field asymmetries as a function of temperature. Here an increase of A$_{\rm TF}$ is observed as $T$ increases, bringing the system from the AFM to the PM state. The evolution of the asymmetry was well fitted using a sigmoid function and the transition temperature is defined as the middle point of the fitting curve T$_N=124.84\pm0.15$~K.

The temperature dependence of the depolarization rate $\lambda_{\rm TF}$ [Fig.~\ref{allasy}(b)] shows the typical cusp at the AFM transition temperature, indicating the onset of spin dynamics in the vicinity of $T_{\rm N}$. It is further noticed that the wTF asymmetry is not completely suppressed even at lowest temperature, which implies that a fraction of the muons ($\sim20\%$) are stopping in a non-magnetic environment. A possible origin of such background is that a fraction of the incoming muons are implanted in the external layers of Mylar tape used to pack the sample. Indeed, the sample we measured in this experiment was rather thin since it could only be synthesized in a small amount, and we used additional layers of Mylar tape to ensure a hermetic sealing, since LiCrTe$_2$ is very sensitive to air and moisture. Finally, it is also noticed that the maximum wTF asymmetry is rather low; approximately 5-10\% is missing. This effect most likely stems from the presence of ferromagnetic (FM) impurities, which are notoriously difficult to avoid in Cr-tellurides \cite{kobayashi2014successive,kobayashi2016competition}. Additionally, the possible formation of muonium during the measurements might also contribute to the asymmetry loss. However, both the background as well as missing fraction, very nicely display one of the powers of the $\mu^+$SR technique, i.e. the capability to discretely separate different volume fractions.

\begin{figure}[ht]
  \begin{center}
    \includegraphics[scale=0.55]{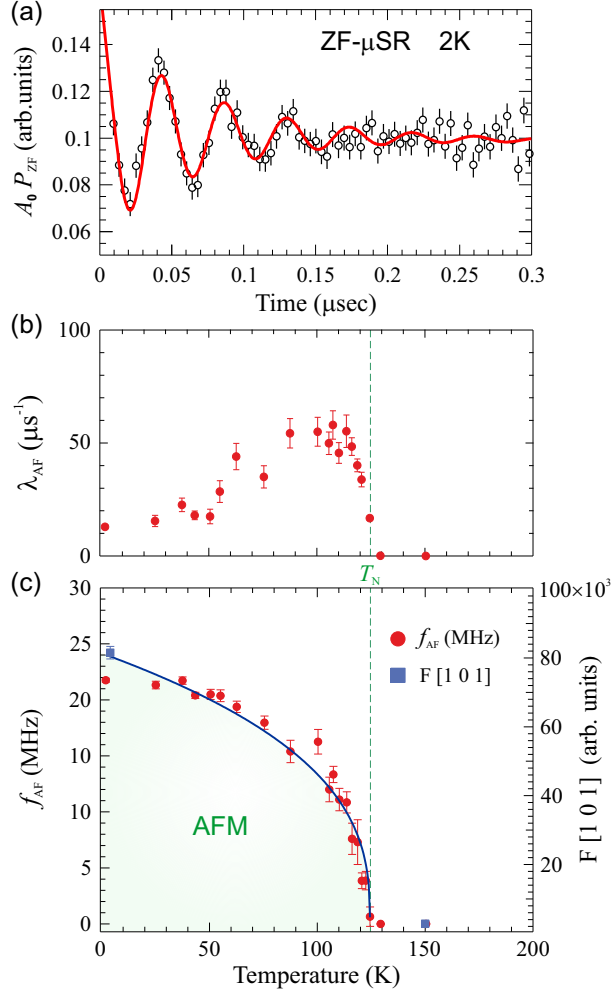}
  \end{center}
  \caption{(a) Early time domain $\mu^+$SR spectra in zero field (ZF) at base temperature. The solid line is a fit to the function in Eq.~\ref{3}. (b) Temperature dependence of the depolarisation rate ($\lambda_{\rm AF}$) of the antiferromagnetic signal. (c) Temperature dependence of the muon spin precession frequency ($f_{\rm AF}$). The blue solid line is a fit of $f_{\rm AF}$(T) to the mean field theory power law in Eq.~\ref{bcs}. The integrated NPD intensity ($F$) of the purely magnetic Bragg peak 1 0 1 at base temperature and above the transition, respectively, are also plotted for comparison (see also text/figures in the "XRD and NPD Results" subsection below).}
  \label{ZFvsfr}
\end{figure}

Another unique feature of $\mu^+$SR is the capability to extract microscopic magnetic properties under true zero external fields. This so-called zero-field (ZF) protocol, was performed for a series of different temperatures, allowing us to observe the evolution of the muon spin rotation and depolarisation. Such properties are directly related to the sample's internal magnetic field distribution (spin order and dynamics) and the acquired ZF $\mu^+$SR time spectrum at base temperature ($T=2$~K) is displayed in Fig.~\ref{ZFvsfr}(a) for the short time domain. Here, the clear oscillations visible in the muon signal are linked to the muons spins' Larmor precession, whose angular frequency ($\omega$) changes according to the local magnetic field (of modulus $B$). The magnetic field intensity distribution in the sample's lattice can be determined through the direct proportionality relation $\omega=\gamma_{\rm \mu}B$ (with the muon´s gyromangetic ratio $\frac{\gamma_{\rm \mu}}{2\pi}=13.55342 [\frac{\rm kHz}{\rm Oe}]$). The time dependence of the ZF muon spin polarization can be described by one or several exponentially relaxing oscillating functions, which are combined to fit the data [solid line in Fig.~\ref{ZFvsfr}(a)]. More precisely, the fit function chosen for the current ZF data is the following:

\begin{eqnarray}
A_0 \, P_{\rm ZF}(t) = A_{\rm AF}\cos(2\pi f_{\rm AF}t + \frac{\pi \phi_{\rm AF}}{180})\cdot{}e^{(-\lambda_{\rm AF} t)} + A_{\rm BG}\cdot{}e^{(-\lambda_{\rm BG} t)} + A_{\rm tail}\cdot{}e^{(-\lambda_{\rm tail} t)}.
\label{3}
\end{eqnarray}

where $A_0$ is the initial asymmetry of the muon decay, $P_{\rm ZF}$ is the muon spin polarization function. The oscillation is well fitted using a single cosine function where $2\pi f_{\rm AF} = \omega$ is the angular frequency of the Larmor precession having the phase $\phi_{\rm AF}$. After correction for conducting the ZF measurements in spin-rotated measurement mode \cite{beveridge1985spin}, we find that $\phi\approx0^{\circ}$ for all recorded temperatures. A zero phase indicate that the magnetic order is commensurate to the crystal lattice. Further, the occurrence of a single frequency in the muon signal is a clear sign for the presence of a single magnetic muon site and, most likely a collinear AFM order.

$\lambda_{\rm AF}$ is the exponential relaxation rate of the cosine function, and $A_{\rm tail} \cdot e^{(-\lambda_{\rm tail} t)}$ is a non oscillatory exponential tail term usually associated to the components of the internal field that are parallel to initial direction of the muon spin polarization. The temperature dependence of the depolarization rate is displayed in Fig.~\ref{ZFvsfr}(b). The increase in the value of $\lambda_{\rm AF}$ around $T_{\rm N}$ reflects the spin dynamics of the localized Cr moments in the system, resulting in a broadening of the field distribution width at the muon sites. This effect is expected in the proximity of a magnetic phase transition as a critical behavior. 


As the temperature increases the spontaneous oscillation of the muon spin polarization (coming from the static internal field associated with the commensurate AFM ordering) is suppressed until only the non oscillatory components of the signal are left. When reaching temperatures above $T_{\rm N}$, the oscillation is completely suppressed. The temperature dependence of the precession frequency of the muon spin in zero field [$f_{\rm AF}(T)$], is proportional to the order parameter of the AFM transition, and it is shown in Fig.~\ref{ZFvsfr}(c). The continuous line is a fit to the mean field theory power law:

\begin{eqnarray}
f(T) \propto f_0 \cdot{} \big(1 - \frac{T}{T_{\rm N}}\big)^{\beta}.
\label{bcs}
\end{eqnarray}

From the fit it is possible to extract the value of the critical exponent as $\beta=0.36\pm0.03$, which suggests a three-dimensional order parameter \cite{yaouanc2011muon}. Consequently, it is found that LiCrTe$_2$ belongs to the 3D Heisenberg AFM universality class. The same fit also yields a value for $T_{\rm N}\approx125$~K, which is fully coherent with the results obtained from the wTF $\mu^+$SR measurements. Such value for $T_{\rm N}$ is very different (higher) than the one previously reported from bulk magnetization measurements, $T_{\rm N}\approx71$~K \cite{kobayashi2016competition}. The reason for such discrepancy is most likely that the lower $T_{\rm N}$ found in bulk magnetic measurements is estimated from data acquired at a high externally applied magnetic field $B=7$~T. Here, the measurements at lower fields ($B=1$~T) only display a very broad feature without the typical AFM kink \cite{kobayashi2016competition}. Such behavior is well known for 2D AFM materials where $T_{\rm N}$ often has to be accurately determined by the use of complementary techniques (see Refs.~\cite{sugiyama2007magnetic,sugiyama2004electron}). The $\mu^+$SR technique allows studies of magnetic properties under very low or even zero applied fields, providing reliable access to spin order/dynamics as well as transition temperature, regardless the dimensionality of the studied system. This ability is due to the high sensitivity of the muon to probe local magnetic environments and weak magnetic moments. Consequently, from our $\mu^+$SR measurements, we are able to extract the true intrinsic magnetic properties of LiCrTe$_2$, revealing a magnetic ground state in the form of a collinear AFM spin order, which is established below T$_N=124.84\pm0.15$~K.

\begin{figure*}[ht]
\centering
  \includegraphics[scale=0.7]{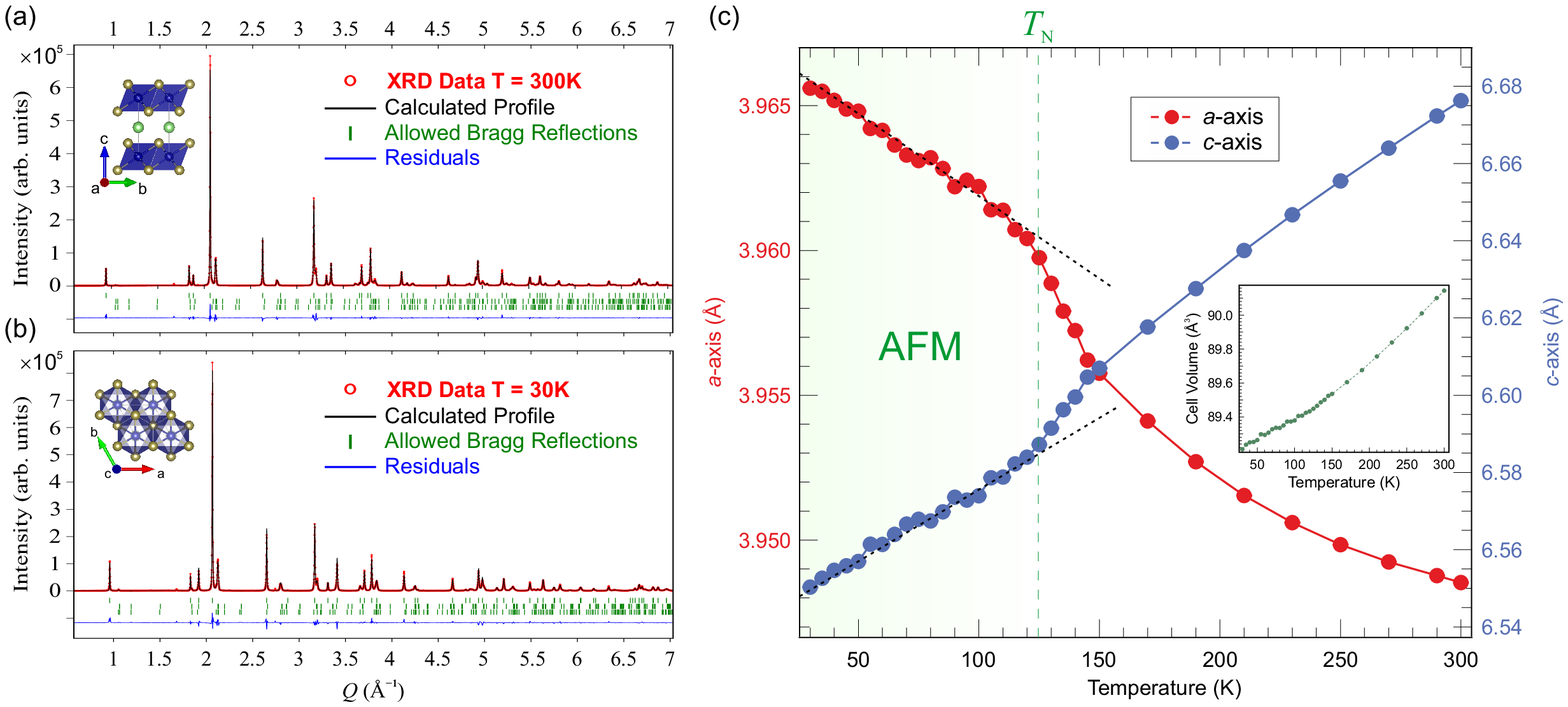}
  \caption{Synchrotron X-ray diffraction (XRD) patterns with the corresponding calculated pattern resulting from Rietveld refinement using spacegroup \textit{P}$\overline{3}$\textit{m}1 (\#164) at (a) $T=300$~K and (b) $T=30$~K. (c) Refined lattice parameters as a function of temperature (see also Table~\ref{xrd_tab}). Dashed and dotted lines are guides to the eye to indicate a change in temperature dependence (slope) at the antiferromagnetic ordering temperature, $T_{\rm N}\approx125$~K. Inset display the temperature dependence of the trigonal cell volume.}
  \label{xrd}
\end{figure*}

\subsection*{XRD and NPD Results}
To robustly determine the spin structure it is first important to have a detailed crystal structure also at low temperature. The crystal structure of polycrystal LiCrTe$_2$, was here deduced from detailed synchrotron XRD data for $T=30-300$~K [see Fig.~\ref{xrd}(a-b)]. As expected from the $\mu^+$SR data, in addition to the main LiCrTe$_2$ phase, two minor Cr$_2$Te$_3$ impurity phases with slightly different lattice parameters were identified and efficiently separated in the refinement. The starting structural parameters for the Cr$_2$Te$_3$ impurity phases were 
retrieved from reference \cite{ipser1983transition}. The main LiCrTe$_2$ phase was modeled with Rietveld refinement with the trigonal spacegroup \textit{P}$\overline{3}$\textit{m}1 (\#164). A substantial improvement of the match between the calculated and experimental patterns was achieved by adding a preferred orientation component along the $\textbf{\textit{c*}}$ reciprocal space direction. This is generally very common (and even expected) in low-dimensional (here 2D) materials.

\begin{figure*}[ht]
\centering
  \includegraphics[scale=0.7]{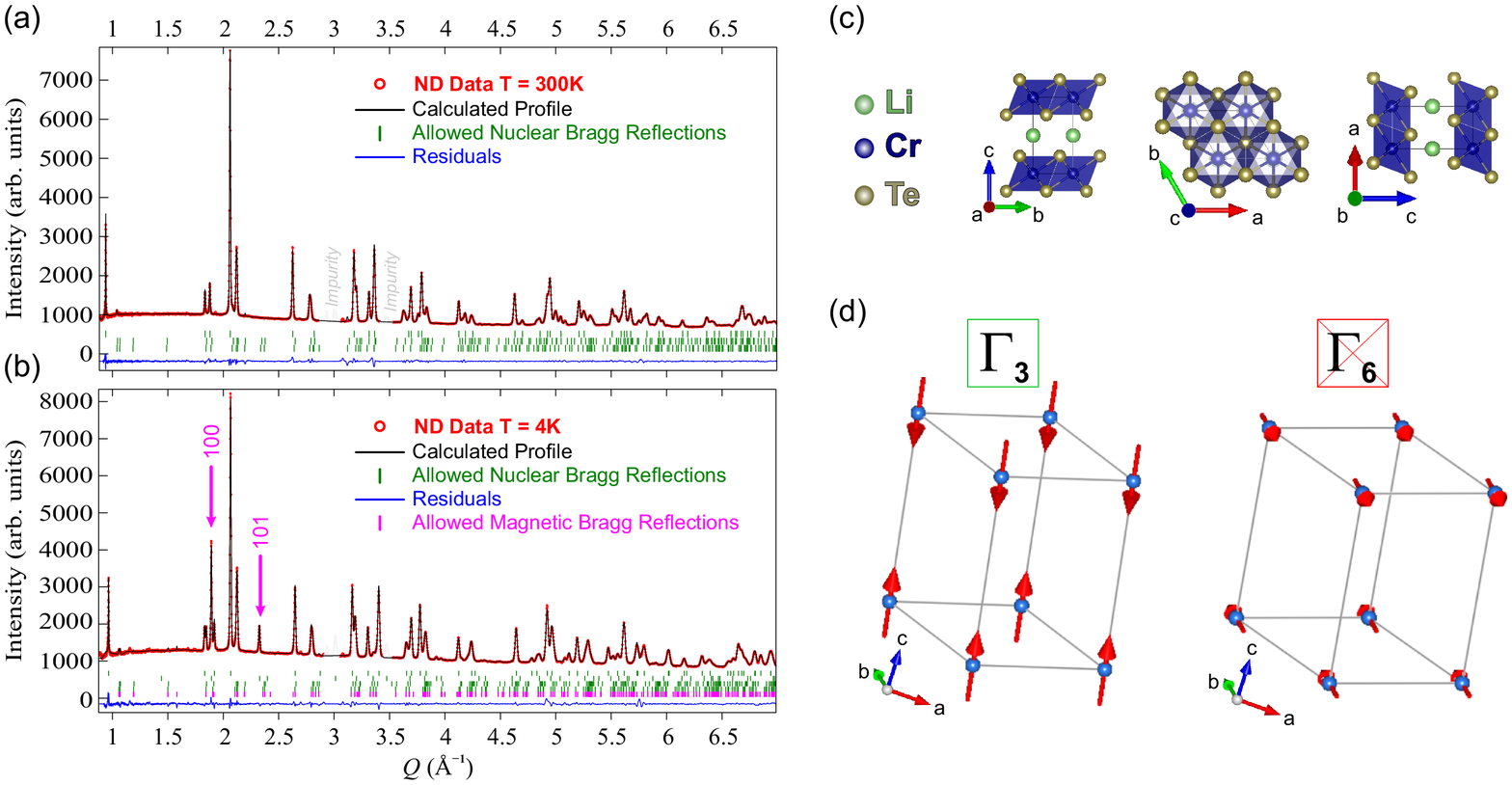}
  \caption{Neutron diffraction pattern with the corresponding calculated pattern resulting from magnetic and structural refinement at (a) $T=300$~K and (b) $T=4$~K. The magenta arrows on the low temperature plot highlights the position of the magnetic peaks. The dotted areas in the pattern around 3 Å$^{-1}$ and 3.5 Å$^{-1}$ are regions in which unknown impurity reflections are located. These regions were excluded from the refinement. (c) Graphic representation of the low temperature crystal structure of LiCrTe$_2$ oriented along the main axes of the unit cell. (d) Graphic representation of the two possible antiferromagnetic spin structures in LiCrTe$_2$. The red arrows represent the ordered Cr moments. Here $\Gamma_3$ clearly gives a much better agreement than $\Gamma_6$. See also Table \ref{table1} for more details.}
  \label{Figure4}
\end{figure*}

The lattice parameters, atomic positions and isotropic thermal displacement parameters $B$ resulting from the Rietveld refinement of the XRD data are summarized in Table~\ref{xrd_tab}. The reliability $R$-factor $R_{B}$ is also reported, the low value of this parameter is indication of the goodness of the model. The refinement of fractional occupancies for both the XRD and NPD data did not substantially improve the agreement between the measured and calculated diffraction patterns, and provided values very close to the ideal stoichiometric sample. The corresponding fully occupied crystallographic sites are reported in Table \ref{xrd_tab} next to the atomic coordinates for each element. The complete temperature dependencies of the lattice parameters are shown in Fig.~\ref{xrd}(c). Here the $c-$axis displays an increase with higher temperatures, compatible with an expected thermal expansion, while the $a-$axis shrinks as the temperature increases. This could be related to the 2D nature of the compound and/or the difference of in-/out-of-plane spin interactions (see below). It is however, clear that the lattice volume (thermal expansion) increases with higher temperatures, as expected [see inset of Fig.~\ref{xrd}(c)]. More interestingly, there is a clear change of slope for both $a-$ and $c-$axis at the magnetic ordering temperature $T_{\rm N}\approx125$~K. Below $T_{\rm N}$ there is a clear linear T-dependence [emphasized by the dotted lines in Fig.~\ref{xrd}(c)], which abruptly changes when the static AFM order vanishes at higher temperatures (note that the nuclear structure/space group remains the same in the entire temperature range). This indicates that there is a strong coupling between the magnetic and lattice degrees of freedom in LiCrTe$_2$. This is in fact very similar to what has been previously reported for the closely related LiCrO$_2$ compound \cite{toth2016electromagnon}. This makes LiCrTe$_2$ a promising material to investigate magnon-phonon interactions, as well as  magnetic excitations, by inelastic X-ray scattering (IXS) (as was done for LiCrO$_2$) \cite{toth2016electromagnon}.

\begin{table}[h!]
\renewcommand{\arraystretch}{1.25}
\small
  \caption{\ Summary of the structural parameters for polycrystal LiCrTe$_2$ refined from synchrotron XRD measurements. A trigonal spacegroup \textit{P}$\overline{3}$\textit{m}1 (\#164) was used for the entire temperature range with $\alpha=\beta=90^{\circ}$ and $\gamma=120^{\circ}$. Also note that refinement of fractional occupancies using both XRD and NPD data resulted in virtually stoichiometric composition (i.e. full occupancies for all sites).}
  \label{xrd_tab}
  \begin{tabular*}{0.48\textwidth}{@{\extracolsep{\fill}}  c c c}
    \hline
                        & 300 K     & 30 K       \\
    \hline
    \textit{a} (\AA)    & 3.94853(1)  & 3.96560(1)  \\
    \textit{c} (\AA)    & 6.67631(3)  & 6.55028(4)  \\
    \hline
    Li ($x$,$y$,$z$)    & (0,0,$\frac{1}{2}$), $1b$  & (0,0,$\frac{1}{2}$), $1b$  \\
    Cr ($x$,$y$,$z$)    & (0,0,0), $1a$              & (0,0,0), $1a$  \\
    Te ($x$,$y$,$z$)   & ($\frac{1}{3}$,$\frac{2}{3}$,0.22675(1)), $2d$          & ($\frac{1}{3}$,$\frac{2}{3}$,0.22616(13)), $2d$ \\
    \hline
    $B_{\rm Li}$ (Å$^2$)   &  2.837(1)          &    1.284(1)            \\
    $B_{\rm Cr}$  (Å$^2$)    &  0.679(1)           &     0.303(1)           \\
    $B_{\rm Te}$   (Å$^2$)   &  0.972(1)           &      0.611(1)          \\
    \hline
    $R_{B}$ (\%) & 2.57       & 5.15        \\
    \hline
  \end{tabular*}
\end{table}

To investigate the sample's low and high temperature nuclear and magnetic structure, neutron powder diffraction (NPD) patterns were collected at the time-of-flight (ToF) instrument iMATERIA, whose design allows to reach a wide range in reciprocal space, while measuring at a single scattering angle by means of large position sensitive detectors. Different detector banks provide different $d$-ranges ($Q$-ranges) with gradually changing resolutions (see the Methods section for more details about the detector banks). 

The neutron diffraction patterns were collected at $T=300$~K and $T=4$~K, from the refinement of the crystal structure we observe, on lowering temperature, a similar (c.f. XRD data) decrease in the value of the \textit{c} axis, suggestive of a shortening of the atomic distances between Cr, Te and Li, accompanied by an increase of the \textit{a} axis, suggestive of an in-plane increase in the distances between Cr atoms located at the corners of the cell.

The magnetic structure refinement was performed on the 4~K diffraction pattern to obtain the detailed spin structure of LiCrTe$_2$ polycrystal. The result of the refinement is displayed in Fig.~\ref{Figure4}(a-b) with a plot of the diffraction patterns (observed and calculated) at high and low temperature, respectively. The magnetic diffraction peaks, $1 0 0$ at $Q=1.9$~Å$^{-1}$ and $1 0 1$ at $Q=2.3$~Å$^{-1}$ are clearly visible in the low temperature diffraction pattern [magenta arrows in Fig.~\ref{Figure4}(b)].

The calculated diffraction pattern contains four phases: a nuclear and a magnetic LiCrTe$_2$ phase plus the two small Cr$_2$Te$_3$ impurity nuclear phases (ferromagnetic, $T_{\rm c}\approx200$~K \cite{Haraldsen_1937,Konno_1993}). The LiCrTe$_2$ magnetic phase, refined by the Rietveld Method, includes a single magnetic Cr atom in its crystallographic site. The scale factor and structural parameters were constrained to be equal to their counterparts in the nuclear LiCrTe$_2$ phase, in order to obtain a correct value for the Cr magnetic moment $\mu_{\rm Cr}$. The magnetic propagation vector $q$ for correct indexing of the temperature dependent magnetic Bragg reflections was determined as $q$ = (0 0 1/2) with the software K-Search. The possible irreducible representations of the propagation vector group $G_k$ were obtained from the paramagnetic parent space group $\textit{P}\overline{3}\textit{m}1$ and the calculated propagation vector $q$ with the software BasIreps. The resulting reducible representation $\Gamma$ was given by the direct sum of two irreducible representations $\Gamma_3$ $\bigoplus$ $\Gamma_6$. Table~\ref{table1} reports the real and imaginary basis vectors of the two representations. Note that $\Gamma_3$ is collinear and $\Gamma_6$ is helical in nature.

\begin{table}[h!]
\renewcommand*{\arraystretch}{1.25}
\small
  \caption{\ Basis vectors of the irreducible representations of the magnetic propagation vector group $G_k$ for the Cr atom in polycrystal LiCrTe$_2$. Note that $\Gamma_3$ is collinear and $\Gamma_6$ is helical (also imaginary component). $\Gamma_3$ clearly gives a better agreement with our NPD data (see main text).}
  \label{table1}
  \begin{tabular*}{0.48\textwidth}{@{\extracolsep{\fill}}  c c c}
    \hline
                        & $\Gamma_3$     & $\Gamma_6$       \\
    \hline

    BsV$_1$(Re)    & (0 0 1)  & (1 0 0)  \\
    BsV$_1$(Im)    & -        & (-0.58 -1.15 0.00) \\
    \hline
    BsV$_2$(Re)    & -       &  (0 1 0)   \\
    BsV$_2$(Im)    & -       & (-1.15 -0.58 0.00) \\
    \hline
  \end{tabular*}
\end{table}

The magnetic structure refinement with IRrep $\Gamma_3$ was able to well capture the temperature dependent magnetic features of the diffraction pattern ($R_{B\rm (mag)}=1.75$\%), while the refinement with IRrep $\Gamma_6$ did not provide a good result ($R_{B\rm (mag)}>50$\%). Therefore the representation $\Gamma_3$, corresponding to the Shubnikov Group $P_c\overline{3}c1$ (\#165.96), was selected to describe the spin structure in polycrystal LiCrTe$_2$. A graphic representation of the two possible IRreps is displayed in Fig.~\ref{Figure4}(d). Here the proposed magnetic structure ($\Gamma_3$) for polycrystal LiCrTe$_2$ is collinear antiferromagnetic with the Cr moments parallel to the $c$ axis. The Cr moments are ferromagnetically coupled within the $ab-$plane while antiferromagnetically coupled along the $c-$axis. The spin only magnetic moment $\mu_{\rm Cr}$ was also determined as 2.36(1) $\mu_B$. This is significantly lower than the the ideal spin-only value of the moment for an isolated Cr$^{3+}$ ($S = 3/2$) ion, which can be calculated as: $\mu_{3/2}$ = 2$\sqrt{S(S+1)}$ = 3.87 $\mu_B$. Further, the value of the effective moment previously reported from bulk magnetic measurements is 3.6 $\mu_B$ \cite{kobayashi2016competition}, which is closer to the ideal value. It is here important to emphasize that in a magnetically ordered bonded system, a reduction in the value of the magnetic moment of a transition metal element with respect to its ideal value might occur \cite{streltsov2016covalent}. Further, the geometric frustration is also well-known to cause staggered magnetic moments and reduction of $\mu_{\rm Cr}$ \cite{Soubeyroux_1979,damay2011magnetoelastic,carlsson2011suppression,Schmidt_2017,Nozaki_2010,Matsubara_2020}. Such reduction is not always possible to discern when estimating the effective moment of a system from the paramagnetic region, which is the case for $\mu_{\rm Cr}$ estimated by bulk magnetic measurements \cite{kobayashi2016competition}. The NPD method is on the other hand able to separate nuclear and magnetic (or even several magnetic) contributions and directly extract the ordered moment inside the magnetic phase even under zero applied field. Consequently, the ordered moment in LiCrTe$_2$ presented in this work ($\mu_{\rm Cr}=2.36(1)~\mu_B$) can be considered reliable. The detailed parameters resulting from the nuclear and magnetic refinements for LiCrTe$_2$ at different temperatures are summarized in Table~\ref{table2} together with their respective $R$-Bragg agreement factors.

\begin{table}[h!]
\renewcommand*{\arraystretch}{1.25}
\small
  \caption{\ Comparison of results from neutron powder diffraction on polycrystal LiCrTe$_2$ at three different temperatures. Nuclear space group \textit{P}$\overline{3}$\textit{m}1 (\#164) and magnetic (Shubnikov) space group $P_c\overline{3}c1$ (\#165.96), i.e. $\Gamma_3$ representation shown in Table~\ref{table1} and Fig.~\ref{Figure4}(d).}
  \label{table2}
  \begin{tabular*}{0.48\textwidth}{@{\extracolsep{\fill}}  c c c c}
    \hline
                        & 300 K & 150K     & 4 K       \\
    \hline

    \textit{a} (\AA)    & 3.950355(3)& 3.95840(6) & 3.967035(5)  \\
    \textit{c} (\AA)    & 6.683390(4)& 6.61042(15) & 6.553415(2)  \\
    $\mu_{\rm Cr}$ ($\mu_B$)& -     &  - & 2.36(1)\\
    $R_{B\rm (nuc)}$ (\%) & 2.63  &   4.5  & 1.99        \\
    $R_{B\rm (mag)}$ (\%)   & -   & -      & 1.75       \\
    \hline
  \end{tabular*}
\end{table}

The observed profiles are in good agreement with the calculated models. The goodness of each model is underlined by the values of the reliability $R$-factors, none of them exceeding a few percent. The magnetic and crystal structure of LiCrTe$_2$ has been determined with a global procedure involving the simultaneous Rietveld refinement of diffraction patterns from the 2 detector banks while keeping the cell parameters, the atomic positions and the basis vectors of the magnetic moment as common parameters.

\section*{Discussion}

The $\Gamma_3$ magnetic structure is compatible with a scenario of super-exchange interaction mechanism, established both inter-planar and intra-planar. Within the $ab-$plane, the ferromagnetic exchange interaction is established among the Cr atoms across the edge-sharing CrTe$_6$ octahedra, with a value of 92.34(3)$^{\circ}$ for the Cr-Te-Cr angle between the two Cr atoms and the shared chalcogen Te. This value is close to 90$^{\circ}$ which, according to the well known Goodenough-Kanamori-Anderson empirical rules for magnetic ordering, favors the ferromagnetic arrangement of the magnetic moments of two cation's through the super-exchange interaction mediated by the anion at the 90$^{\circ}$ point between them \cite{goodenough1955theory,kanamori1959superexchange,anderson1950antiferromagnetism}. Along the $c-$axis, the out-of-plane AFM arrangement of the Cr magnetic moments is mediated by super-exchange interactions through two Te atoms, across the path Cr-Te-Te-Cr. This process involves two anions, therefore it cannot straightforwardly be described in terms of the simple Goodenough-Kanamori-Anderson rules, and gives rise to a much weaker inter-layer coupling, as also pointed out in previous studies \cite{kobayashi2016competition}. Finally, we discuss the possibility of a competition between a direct Cr-Cr exchange and the Cr-Te-Cr super-exchange mechanism here suggested for the adjacent Cr atoms in the $ab-$plane. Here, it should be noted that, since Cr$^{3+}$ ions have an electronic configuration with half filled $t_{2g}$ orbitals, the direct exchange would lead to an AFM coupling of the Cr spins within the $ab-$layer. The results obtained in this work show a clear predominance of the FM super-exchange interaction over the AFM direct exchange within the Cr layers of LiCrTe$_2$ (i.e., the $ab$-plane). Therefore, a competition between the two coupling mechanisms does not seem to take place in this material. This is consistent with the observations of Rosenberg $et$ $al.$ \cite{rosenberg1982magnetic}, who identified a minimum critical Cr-Cr distance of $\approx$ 3.6 Å in the Cr layers of ternary chromium chalcogenides, for the occurrence of FM coupling. Since the Cr-Cr distance below $T_{\rm N}$ in LiCrTe$_2$ is 3.967 Å $>$ 3.6 Å, it is very reasonable that the FM mechanism is dominant. This is also consistent with the increase in $a$-axis observed at low temperature in the XRD data, which seems to occur to better accomodate the in-plane FM interaction.\\

To summarize the outcomes of this work, we present the first detailed study of the magnetic ground state in LiCrTe$_2$, where we successfully reveal the true intrinsic spin properties, fully disentangled from any interference from ferromagnetic impurities or external fields. The structural and magnetic properties of polycrystal LiCrTe$_2$ have been investigated by state-of-the-art high resolution large-scale experimental techniques. From the neutron diffraction measurements, the magnetic structure of LiCrTe$_2$ was determined to be collinear antiferromagnetic, with the Cr moments oriented in ferromagnetic layers within the $ab-$plane, and antiferromagnetically coupled along the $c$-axis. The spin only moment for the Cr atom $\mu_{\rm Cr}$ was determined as 2.36(1) $\mu_B$ and the super-exchange interaction was identified as coupling mechanism for the Cr spin ordering. From $\mu^+$SR investigations the profile of the order parameter for the magnetic phase transition in LiCrTe$_2$ was outlined, allowing us to categorize this material as a 3D Heisenberg antiferromagnet with magnetic transition temperature T$_N=124.84\pm0.15$~K. Combining our magnetic results with high-resolution synchrotron X-ray diffraction, we find a strong spin-lattice coupling that could yield future interesting studies of phonon-magnon interactions.

\section*{\label{methods}Methods}

Polycrystalline samples of LiCrTe$_2$ were prepared by means of conventional solid state synthesis. Thereby, Li, Cr, and Te elements were reacted directly. Further details on the sample preparation can be found in Ref.~\cite{kobayashi2016competition}.

The $\mu^+$SR spectra have been acquired, as a function of temperature, at the multi purpose muon beam-line M20, at the Canada's particle accelerator centre TRIUMF. About 200 mg of powder sample was packed in a $5\times5$~mm$^2$ envelope made out of Aluminum-coated Mylar tape (0.05~mm thickness). The envelope was hermetically sealed using a second layer of Mylar tape. The envelope was attached to a low background sample holder inserted in a helium exchange gas cryostat ($T=2-300$~K).

The neutron diffraction measurements were performed at the time of flight powder diffractometer iMATERIA \cite{ishigaki2009ibaraki} at the high intensity proton accelerator facility J-PARC (Japan). The powder sample ($m\approx400$~mg) was introduced into a cylindrical vanadium cell with a diameter \o~=~5~mm and sealed using an indium wire together with an aluminium cap and screws. The cell was mounted onto a closed cycle refrigerator (CCR) to reach temperatures $T=2-300$~K. In iMATERIA the backward detector bank (BS), allows a $d$-range from 0.181 Å up to 5.09 Å with a resolution $\Delta d / d$ = 0.16$\%$, the 90-degree detector bank (SE) allows a $d$-range from 0.255 Å up to 7.2 Å with a resolution $\Delta d / d$ = 0.5$\%$, while the low angle (LA35) detector bank has a $d$-range from 0.25 Å up to 40 Å. These features make the higher angle banks very suitable for detailed structural characterization while the low angle banks are ideal to unequivocally identify any magnetic Bragg peak, which usually appear in the high-$d$ (low-$Q$) range. In the specific case under investigation only the BS and SE banks were used in the refinement since no additional magnetic Bragg peaks were observed in the lower angle banks.

The synchrotron XRD measurements were collected at the BL02B2 instrument in SPring-8, synchrotron radiation facility in Japan. XRD profiles were collected using a high-resolution one-dimensional solid-state detector (MYTHEN) \cite{kawaguchi2017high}. The powdered samples were sealed in a borosilicate glass capillary with a diameter of 0.3 mm. Temperature control was conducted using a He gas stream device, in a temperature range $T=30-300$~K. The data collection was carried out with a wavelength $\lambda=0.413627$~Å for the incoming photon beam and 2 minutes acquisition time.

Finally, due to high sensitivity of the sample to air and moisture, the entire sample mounting procedures for all the measurements were carried out in glove-boxes under Ar or He atmosphere, in order to avoid sample degradation.

The crystal and magnetic structure determination was carried out with the FullProf software suite \cite{rodriguez1993recent}. All images involving crystal structure were made with the VESTA software \cite{momma}, the parameter fitting has been performed with the software IgorPro \cite{igor} and the $\mu^+$SR data were fitted using the software package \textit{musrfit} \cite{musrfit}.

\bibliography{Refs}

\section*{Acknowledgements}

The $\mu^+$SR measurements were performed at the instrument M20 of the muon source TRIUMF (beamtime proposal: M1673). The XRD measurements were performed at the instrument BL02B2 of the synchrotron facility SPring-8 (beamtime proposal: 2019B1792). The NPD measurements were performed at the instrument iMATERIA of the neutron spallation source J-PARC (beamtime proposal: 2019A0330).
The authors wish to thank Dr. Hiroshi Nozaki for his support during the $\mu^+$SR experiment along with the staff of TRIUMF, SPring-8 and J-PARC for the invaluable help in the experimental measurements. All the figures of this paper have been drawn by E.N. and M.M.
This research is funded by the Swedish Foundation for Strategic Research (SSF) within the Swedish national graduate school in neutron scattering (SwedNess), as well as the Swedish Research Council VR (Dnr. 2021-06157 and Dnr. 2017-05078), and the Carl Tryggers Foundation for Scientific Research (CTS-18:272). J.S. is supported by the Japan Society for the Promotion Science (JSPS) KAKENHI Grant No. JP18H01863 and JP20K21149. Y.S. and O.K.F. are funded by the Chalmers Area of Advance - Materials Science. S.K. and K.Y. are supported by JSPS KAKENHI Grant No.18KK0150. The work at UZH and UNiGe was supported by the Swiss National Science Foundation under Grant No. PCEFP2-194183.

\section*{Author contributions statement}

E.N., J.S., and M.M. conceived the experiments. E.N., O.K.F., N.M., F.M., A.Z., I.U., Y.S., J.S. and M.M. conducted the experiments. E.N, J.S., N.M., O.K.F. and M.M. analyzed the results. The samples were synthesized by S.K. and K.Y., as well as by C.W. and F.O.v.R.; they also conducted the initial sample characterizations. E.N. created the first draft, and all co-authors reviewed and revised the manuscript. 


\section*{Additional information}

\textbf{Competing interests} The authors declare no competing interests.  

The corresponding author is responsible for submitting a \href{http://www.nature.com/srep/policies/index.html#competing}{competing interests statement} on behalf of all authors of the paper. 

\end{document}